\documentclass[5p,twocolumn]{elsarticle}

\usepackage{epsf,epsfig,amsmath,amssymb}
\usepackage{color}

\newsavebox{\ns}
\newsavebox{\dbrane}

\def\be{\begin{equation}}
\def\ee{\end{equation}}
\def\bea{\begin{eqnarray}}
\def\eea{\end{eqnarray}}

\def\Dslash{\,\,{\raise.15ex\hbox{/}\mkern-12mu D}}
\def\Dbarslash{\,\,{\raise.15ex\hbox{/}\mkern-12mu {\bar D}}}
\def\delslash{\,\,{\raise.15ex\hbox{/}\mkern-9mu \partial}}
\def\delbarslash{\,\,{\raise.15ex\hbox{/}\mkern-9mu {\bar\partial}}}
\def\pslash{\,\,{\raise.15ex\hbox{/}\mkern-9mu p}}
\def\calDslash{\,\,{\raise.15ex\hbox{/}\mkern-12mu {\cal D}}}

\newcommand{\vol}{\mbox{vol}}

\newcommand{\nn}{\nonumber \\}

\def\w{\wedge}

\allowdisplaybreaks

\begin{document}


\title{\vspace*{-35pt} \flushright{\small{KIAS. P11034 \quad AEI-2011-036}}
\\[15pt]
\center{Consistent reductions from $D=11$ beyond Sasaki-Einstein}}

\author[rvt]{Eoin \'O Colg\'ain}
\ead{eoin@kias.re.kr}
\address[rvt]{Korea Institute for Advanced Study, Seoul 130-722, Korea}
\author[focal]{Oscar Varela}
\ead{oscar.varela@aei.mpg.de}
\address[focal]{AEI, Max-Planck-Institut f\"ur Gravitationsphysik, Am M\"uhlenberg 1, D-14476 Potsdam, Germany}

\begin{abstract}

The most general class of warped $AdS_5 \times_w M_6$ supersymmetric solutions in $D=11$ supergravity permits a consistent truncation to $D=5$ $N=2$ minimal gauged supergravity. Here we extend this truncation, for a particular subclass of $M_6$ manifolds, to $D=5$ $N=4$ gauged supergravity coupled to two vector multiplets. We obtain the reduction ansatz by T-duality of a recently discussed type IIB truncation on a generic Sasaki-Einstein five-fold, which becomes non-trivial in $D=11$ and displays non-standard features due to the $G$-structure of the internal $M_6$. Using this truncation, we construct two new classes of warped and unwarped non-supersymmetric $AdS_5$ M-theory backgrounds. We also comment on possible extensions of the reduction ansatz to the general class of $M_6$ geometries.

\end{abstract}

\maketitle


\section{Introduction} \label{Introduction}

The lower dimensional effective theories that result from string or M-theory compactification provide a very useful arena for phenomenological, cosmological or holographic studies. In this context, the requirement that the lower-dimensional effective theory be a {\it consistent truncation} of the higher-dimensional one, namely, that all solutions of the lower-dimensional equations of motion lift to higher-dimensional solutions, has a variety of formal and practical applications. For example, in the absence of a clear separation of scales in flux compactifications, consistent truncations at least provide a well defined mechanism to retain a finite number of lower-dimensional fields. By their very definition, they prove valuable tools in the construction of higher-dimensional supergravity backgrounds and, in AdS/CFT, have been recently found useful to embed gravitational toy models into UV-safe string theory.

Consistent truncations on $G$-structure manifolds are relatively easy to construct whenever there is some symmetry principle that  selects the set of expansion forms and determines the effective fields to be kept in the truncation. This is the case of the consistent truncation of M-theory on Sasaki-Einstein seven-folds ($SE_7$) of \cite{Gauntlett:2009zw}. In this case, $SE_7$ comes equipped with an $SU(3)$ structure that contains only constant, $SU(3)$-singlet torsion classes. The invariant forms $(\eta , J, \Omega)$ that define the $SU(3)$-structure can be safely used as expansion forms while the constant, singlet torsion classes guarantee the consistency of the truncation. Essentially the same mechanism was put at work in \cite{KashaniPoor:2007tr} in the context of (massive) type IIA truncations on nearly-K\"ahler ($NK_6$) six-folds: again, $NK_6$ is endowed, by definition, with an $SU(3)$-structure with only a singlet torsion class, $W_1$. Similar truncations are possible for homogeneous internal spaces $G_0/H_0$ enjoying a $G_0$--invariant $G$-structure, see \cite{Micu:2006ey}--\cite{Bena:2010pr}. Now, $G_0$ invariance guarantees consistency in {\it e.g.} the truncations of type IIA on $SU(3)$-structure cosets of \cite{Cassani:2009ck}, in spite of the internal spaces being half-flat (and, thus, having a non-$SU(3)$-singlet class $W_2$ in addition to $W_1$) at generic points in moduli space \cite{Koerber:2008rx}.

This situation changes dramatically when no obvious symmetry principle can be invoked in order to select the expansion forms. Remarkable examples of consistent truncations on $G$-structure manifolds that do not seem to rely on symmetry arguments do nevertheless still exist, the most prominent ones being the maximally supersymmetric truncations of $D=11$ supergravity on (identity-structure) spheres down to $D=4$ $N=8$ $SO(8)$--gauged supergravity \cite{de Wit:1986iy}, or $D=7$ $N=2$ $SO(5)$--gauged supergravity \cite{Nastase:1999cb}. More generally, consistent truncations on non-trivial $G$-structure manifolds down to pure (ie, with no matter couplings) supergravity with reduced supersymmetry have been conjectured (and shown in some cases) to always exist \cite{Gauntlett:2007ma} with, again, no clear symmetry principles at work. A specific example of this situation is provided \cite{m6red} by the consistent truncation of $D=11$ supergravity on $M_6$ down to minimal $D=5$ gauged supergravity, where $M_6$ is the class of $SU(2)$-structure manifolds, discussed by Gauntlett, Martelli, Sparks and Waldram (GMSW) in \cite{ads5m6}, whose torsion classes are defined by the requirement that the warped product $AdS_5 \times_w M_6$ be a supersymmetric ($N=2$) solution of $D=11$ supergravity.

The consistent truncation of \cite{m6red} (as well as those of \cite{Gauntlett:2007ma, Gauntlett:2007sm}) reveals an intriguing generosity of the class of manifolds $M_6$ of \cite{ads5m6}: not just that the (only) $AdS_5$ vacuum of minimal $D=5$ gauged supergravity uplifts to $D=11$ on such $M_6$ (which was in fact designed in \cite{ads5m6} to do precisely this job), but in fact {\it any} solution of the minimal $D=5$ theory can be uplifted to $D=11$ on {\it the same} $M_6$. This situation prompts the question of whether $M_6$ allows for the uplift of even more general $D=5$ configurations, namely, whether it admits consistent truncations including $N=2$ (or even $N=4$, on account of its $SU(2)$-structure) matter couplings. In this note we will show that this does happen for a particular subclass of GMSW geometries, even in the absence of the symmetry principles that allow for this enlargement to occur \cite{Gauntlett:2009zw,Cassani:2010uw, oscar1, Liu:2010sa} (see also \cite{Skenderis:2010vz}) in {\it e.g.} Sasaki-Einstein truncations \cite{buchel,Gauntlett:2007ma}. We will focus on the bosonic sector of supergravity. For recent results on the fermionic extensions in the Sasaki-Einstein truncation context, see \cite{Bah:2010yt}--\cite{Liu:2011dw}.

With no guiding symmetry at hand, we resort to indirect methods to construct the relevant Kaluza-Klein (KK) ansatze. In section \ref{Mtheory}, we present a consistent truncation of $D=11$ supergravity on a particular class of GMSW manifolds down to the same $D=5$ $N=4$ matter-coupled supergravity (briefly reviewed in section \ref{Background}) that arises from type IIB on a generic Sasaki-Einstein five-fold $SE_5$ \cite{Cassani:2010uw, oscar1}. We construct the $D=11$ reduction via T-duality from type IIB, and stress that the former, unlike the latter, no longer follows from symmetry principles. Using this result, in section \ref{solutions} we construct two new classes of non-supersymmetric $AdS_5$ solutions in $D=11$. The reduction ansatz of section \ref{Mtheory} does not easily extend to the most general class of GMSW $M_6$ manifolds and, in section \ref{Remarks}, we comment on some of the difficulties to write down a consistent ansatz involving the defining forms of the $SU(2)$-structure on general $M_6$. Section \ref{Conclusions} concludes.

\section{Background} \label{Background}
In this section, we review the GMSW class of solutions, the reduction to minimal $D=5$ $N=2$ supergravity, and the $D=5$ $N=4$  $SE_5$ reduction of type IIB supergravity. We also address how the latter may be tailored to the explicit $Y^{p,q}$ metrics. This will be useful to construct a reduction to the $D=5$ $N=4$ theory from $D=11$.

\subsection{$AdS_5 \times_{w} M_6$ geometries and $N=2$ reduction} \label{geometries}
We begin by summarising the GMSW solutions \cite{ads5m6}. The $D=11$ geometry is a warped product of $AdS_5$ and an internal Riemannian manifold $M_6$,
\be \label{GMSWmetric11}
ds^2 = e^{2 \lambda}  [ ds^2(AdS_5) + ds^2(M_6)],
\ee
where the warping $\lambda$ depends only on the coordinates on $M_6$. Supersymmetry imposes an $SU(2)$-structure on $M_6$, defined by two real one-forms $K^1$, $K^2$, a real two-form $J$ and a complex two-form $\Omega$. The metric on $M_6$ reads
\bea \label{GMSWmetric}
 ds^2(M_6) &=& e^{-6 \lambda} ds^2(M_4) + (K^1)^2 + (K^2)^2 ,
\eea
where, introducing coordinates $y, \psi$, one can write $K^1 = e^{-3 \lambda} \sec \zeta dy$, $K^2 = \tfrac{1}{3} \cos \zeta (d \psi + \rho)$, $M_4$ is a K\"ahler manifold, $\psi$ parametrises the R-symmetry direction (with $K^2$ related to the corresponding Killing vector), and the function $\zeta$ and the connection $\rho = \rho_i dx^i$ depend on the coordinate $y$ and the coordinates on $M_4$. The warp factor $\lambda$ and  $\zeta$ are also related to $y$ via $2 y = e^{3 \lambda} \sin \zeta$. We will not need the expression for the background four-form.

When $M_6$ is complex, we have the following simplifications: $d_4 \zeta = d_4 \lambda = \partial_y \rho = 0$,
in which case, $\rho$ becomes the canonical connection on the K\"{a}hler manifold $M_4$. The supersymmetry conditions may then be integrated leading to a class of geometries which are topologically $S^2 $-bundles, parameterised by $(y, \psi)$,  over smooth K\"{a}hler bases $M_4$ which are either four-dimensional K\"ahler-Einstein spaces, $M_4 = KE_4$, or products of constant curvature Riemann surfaces, $M_4 = \mathcal{C}_1 \times \mathcal{C}_2$. Further details, as well as the explicit $SU(2)$ torsion conditions and the various expressions for and $e^{6 \lambda}, \cos^2 \zeta$ and $J$ characterising individual solutions may be found in \cite{ads5m6}. For future reference, here we would just like to mention that, for the specific class of solutions with product base, including $M_4 = S^2 \times T^2$, the non-zero modules characterising the $SU(2)$-structure of $M_6$ are, in the notation of \cite{Dall'Agata:2003ir}, the singlets $S_2$, $S_3$, $S_7$, $S_8$, and the two-forms $T_1$, $T_2$. Further, the singlets are non-trivial functions of $\lambda$ and $\zeta$.

$D=11$ supergravity admits a consistent truncation on the GMSW class of geometries to $D=5$ $N=2$ minimal gauged supergravity \cite{m6red}, the field content of which comprises the metric $ds_5^2$ and the graviphoton, $A_1$. The KK ansatz for the metric is obtained from (\ref{GMSWmetric11}), (\ref{GMSWmetric}) by replacing $ds^2(AdS_5)$ with $ds_5^2$, and $K^2$ with $e^{6} = K^2 + \tfrac{1}{3} \cos \zeta A_1$. The reduction ansatz for the four-form reads
\bea
\label{fluxred}
\tilde{G}^{(4)} &=& {G}^{(4)}_0 + \frac{e^{3 \lambda}}{3} \left( -\sin \zeta J + K^1 e^6 \right) \wedge F_2 \nn &+& \frac{1}{3} e^{3 \lambda} \cos \zeta K^1 \wedge * F_2,
\eea
where $F_2 = d A_1$, $*$ is the Hodge dual with respect to $ds_5^2$, and ${G}^{(4)}_0$ is the background four-form with again $K^2$ replaced with $e^6$. Substitution into the $D=11$ equations of motion shows the consistency of this ansatz.

\subsection{Type IIB $SE_5$ reduction} \label{SE5reduction}

Type IIB supergravity on a generic $SE_5$ manifold reduces to $D=5$ $N=4$ gauged supergravity coupled to two vector multiplets \cite{Cassani:2010uw, oscar1}. The KK ansatz for the metric naturally employs the structure of a $U(1)$ fibration over a K\"ahler-Einstein four-dimensional base ($KE_4$) of $SE_5$. Following the IIB Einstein frame conventions of \cite{oscar1}, it reads
{\setlength\arraycolsep{-5pt}
\bea\label{KKmet}
&& ds^2_{IIB}=e^{-\frac{2}{3}(4U+V)}ds^2_{5}+e^{2U}ds^2(\textrm{$KE_4$}) + e^{2V}(\eta+A_1)^2. 
\eea
}Here $ds^2_5$ is an arbitrary (Einstein frame) metric on five-dimensional space-time, $\eta$ is the dual one-form of the Reeb vector of $SE_5$,
$U$ and $V$ are scalar fields and $A_1$ is a one-form defined on the external five-dimensional space.

The ansatz for the form field strengths is given by
{\setlength\arraycolsep{-3pt}
\begin{eqnarray} \label{KKforms}
&& F_{(5)} = 4 e^{-\frac{8}{3}(4U+V)+Z} \textrm{vol}^{(E)}_5 +e^{-\frac{4}{3}(U+V)} * K_2 \wedge J \nn && \quad + K_1 \wedge J \wedge J  + 2e^Z J \wedge J  \wedge (\eta+A_1) \nn && \quad 
-  \left[ 2e^{-8U} *K_1   - K_2 \wedge J \right] \wedge (\eta+A_1)\nn && \quad 
  + \left[ e^{-\frac{4}{3}(U+V)} *L_2 \wedge \Omega 
  +L_2 \wedge \Omega \wedge (\eta+A_1) +c.c. \right],
\nonumber \\
&& F_{(3)} = G_3 +G_2 \wedge (\eta+A_1) +G_1 \wedge J \nn && \quad 
+ \left[ N_1 \wedge \Omega +N_0 \Omega \wedge (\eta+A_1) +c.c. \right],
\nonumber \\
&& H_{(3)} = H_3 +H_2 \wedge (\eta+A_1) +H_1 \wedge J \nn  && \quad 
+ \left[ M_1 \wedge \Omega +M_0 \Omega \wedge (\eta+A_1) +c.c. \right], \nn 
&&  C_{(0)} = a, \quad \Phi = \phi.
\end{eqnarray}
}Now, $(J, \Omega, \eta)$ characterise the $SU(2)$ structure on $SE_5$. Further details and expressions for the above spacetime forms $K_i, L_i, G_i, H_i , N_i$ and $M_i$ in terms of potentials may be found in \cite{oscar1}. Here we only note that the $D=5$ degrees of freedom include, besides the metric, eleven scalars (seven real, $U, V, \phi, a, h, b, c$,  and two complex, $\xi, \chi$) parametrising the $N=4$ moduli space $SO(1,1) \times \frac{SO(5,2)}{SO(5) \times SO(2)}$; two real two-forms $B_2, C_2$; one complex two-form $L_2$; and four real vectors $A_1, B_1, C_1, E_1$. The latter are gauge fields of $U(1) \times H_3$, where $U(1)$ corresponds to the R-symmetry, and $H_3$ is the three-dimensional Heisenberg group. Note that $K$ appears with superscripts for GMSW vectors and also with subscripts where it corresponds to a field in the reduction.

We now particularise the reduction ansatz (\ref{KKmet}), (\ref{KKforms}) to $SE_5 = Y^{p,q}$. We find it  convenient to use global coordinates $(\theta, \varphi, y, \alpha, \psi)$ (see \cite{ypq}), rather than the local ones suitable to display the Reeb fibration of $Y^{p,q}$. In global coordinates, the $SU(2)$-structure forms on $Y^{p,q}$ read
{\setlength\arraycolsep{1pt}
\bea \label{SU2str}
\eta &=& \frac{1}{3} \left[ (1-\tilde{c}y) \sigma^3 - 6 y d \alpha \right], \nn
J &=& \frac{1}{6} \left[ - dy \w (\tilde{c} \sigma^3 + 6 d \alpha) + (1-\tilde{c}y) \sigma^1 \wedge \sigma^{2}  \right], \\
\Omega &=&  \sqrt{\frac{1-\tilde{c} y}{6 w q}} (\sigma^2 - i\sigma^1) \w [dy - i\frac{wq}{6} (6 d \alpha + \tilde{c} \sigma^3)], \nonumber 
\eea
}where $\sigma^i$, $i=1,2,3$, are $SU(2)$ left-invariant forms (not directly related to the $SU(2)$-structure). In these coordinates, $\sigma^2 - i\sigma^1 = e^{i\psi}(d\theta + i \sin \theta d \varphi)$, $\sigma^3 = d\psi-\cos\theta d \varphi$.  In addition, $w$ and $q$ are functions of $y$
{\setlength\arraycolsep{-3pt}
\bea \label{funwq}
&& w \equiv e^{6\lambda} = \frac{2(\tilde{a}-y^2)}{1-\tilde{c} y}, \quad 
q  \equiv \cos^2 \zeta = \frac{\tilde{a}-3 y^2 + 2 \tilde{c} y^3}{\tilde{a}-y^2},
\eea
}where $\tilde{a}$ and $\tilde{c}$ are constants. It may be easily verified that these forms satisfy the $SE_5$, $SU(2)$-singlet torsion conditions, namely $d \eta = 2 J$ and $ d \Omega = 3 i \eta \w \Omega$.

In these coordinates, the ansatz (\ref{KKmet}) for $SE_5 = Y^{p,q}$ is
{\setlength\arraycolsep{-3pt}
\bea \label{KKmetYpq}
&& ds^{2}_{IIB} = e^{-\frac{2}{3}(4U+V)}ds^2_{(E)} 
+e^{2 U}  \frac{1- \tilde{c}y}{6} ds^2(S^2) + e^{2 U} \frac{dy^2}{w q}  \nn && \qquad + \frac{e^{2 U + 2 V} wq}{9 \Delta} (\hat{\sigma}^3)^2 + \Delta [d \alpha + \tilde{A}_{(1)} ]^2 ,
\eea
}where $ds^2(S^2) = (\sigma^1)^2 + (\sigma^2)^2 = d\theta^2 + \sin^2\theta d\varphi^2$, $\hat{\sigma}^3 = \sigma^3 + 3 A_1$, we have defined 
\begin{equation} \label{eq:delta}
\Delta = e^{2 U} wq + 4 y^2 e^{2 V},
\end{equation}
(note that $\Delta = e^{6\lambda}$ for $U=V=0$) and
\begin{equation} \label{connection1}
\tilde{A}_{(1)} = - \frac{\tilde{c}}{2} A_1 + \frac{1}{6 \Delta} [e^{2 U} \tilde{c} wq - e^{2 V} 4 y (1-\tilde{c} y)] \hat{\sigma}^3
\end{equation}
fibers the $S^1$ with coordinate $\alpha$ both over spacetime and over the squashed $S^3$ parametrised by $\sigma^i$.
Note that $A_1$ in (\ref{fluxred}) is 3 times $A_1$ in $\hat{\sigma}^3$.
%
%

Finally, the KK ansatz for the IIB form field strengths is given by (\ref{KKforms}) with (\ref{SU2str}). In particular, the $H_{(3)}$ field strength may be integrated to give the $B$-field 
{\setlength\arraycolsep{-4.5pt}
\bea
&& B_{(2)} = B_2 -\frac{1}{6} Db \wedge  [(1-\tilde{c} y) \hat{\sigma}^3 - 6 y D 
\alpha ]  + [ \xi \Omega + c.c.],
\eea
}where $Db \equiv db -2B_1$ and $D\alpha \equiv d\alpha -\frac{\tilde{c}}{2} A_1$.

\section{$D=5$ $N=4$ gauged supergravity from $D=11$} \label{Mtheory}

The $Y^{p,q}$ manifolds were originally discovered from the GMSW geometry corresponding to complex $M_6$, with $M_4= S^2 \times T^2$ in (\ref{GMSWmetric}),  by reducing and T-dualising along $T^2$ \cite{ads5m6}. We will now reverse this process, in order to uplift the KK ansatz (\ref{KKmetYpq}), (\ref{KKforms}) to $D=11$. Since this process maps solutions to solutions, the consistency of the ansatz thus obtained for the reduction of $D=11$ supergravity to the full $D=5$ $N=4$ gauged supergravity of \cite{Cassani:2010uw, oscar1} is guaranteed. Below we will argue, however, that the reduction becomes non-trivial from a $D=11$ point of view. 

We will first deal with the NS sector. After T-dualising on the $\alpha$-direction (using consistent conventions e.g. \cite{hassan}), the (Einstein frame) metric becomes 
{\setlength\arraycolsep{-4pt}
\bea \label{KKmetYpqIIA}
&& ds^{2}_{IIA} = e^{\frac{\phi}{8}} \Delta^{\frac{1}{4}} \biggl[  e^{-\frac{2}{3}(4U+V)}ds^2_{5} 
+e^{2 U}  \frac{1- \tilde{c}y}{6} ds^2(S^2) \nn &&  + e^{2 U} \frac{dy^2}{w q} +  \frac{e^{2 U + 2 V} wq}{9 \Delta} (\hat{\sigma}^3)^2 \biggl] + e^{-\frac{7}{8}\phi} \Delta^{-\frac{3}{4}} [d \alpha + \tilde{B}_{(1)}]^2 \
\eea
}where now
{\setlength\arraycolsep{-3pt}
\begin{eqnarray} \label{connection2}
&& \tilde{B}_{(1)} = y Db - \left( \frac{i}{\sqrt{6}} \xi \sqrt{wq(1-\tilde{c}y})(\sigma^2 - i \sigma^1) + c.c. \right)
\end{eqnarray}
}still fibers $S^1$ over spacetime and over the squashed $S^3$, but now the fibration is over the $S^2$ base of the latter. The IIA dilaton is $e^{2\tilde{\phi}} = e^{\frac{3}{2}\phi} \Delta^{-1}$ and the transformed $B$-field is
{\setlength\arraycolsep{0pt}
\bea
&& \tilde{B}_{(2)} = B_2 - \frac{\tilde{c}}{2} A_1 \wedge d \alpha  - \frac{e^{2 U} wq}{6 \Delta} Db \w \hat{\sigma}^3  \nn && \ + \frac{1}{6 \Delta} [e^{2 U} \tilde{c} wq - e^{2 V} 4 y (1-\tilde{c} y)] \hat{\sigma}^3 \wedge d \alpha    \\ && \ +
\left[  \xi \sqrt{\frac{1-\tilde{c} y}{6 w q}} ( \sigma^1 - i \sigma^1) \wedge  ( dy  - i \frac{2 wq y e^{2 V} }{3 \Delta} \hat{\sigma}^3 ) +c.c. \right] . \nonumber
\eea
}

To find the $D=11$ metric, we need to identify the RR one-form potential. This can be calculated to be
{\setlength\arraycolsep{0pt}
\bea
&& \tilde{C}_{(1)} = a d \alpha - y (Dc - a Db) \nn && \quad + \left( \frac{i}{\sqrt{6}} (\chi - a \xi) 
\sqrt{wq(1-\tilde{c} y)}(\sigma^2 - i \sigma^1)
+c.c. \right)
\eea
}where $Dc \equiv dc - 2C_1$.
Uplifting now this type IIA solution, we eventually find the following $D=11$ metric:
{\setlength\arraycolsep{-2pt}
\bea
\label{upliftmet}
&& ds^2_{11} = \Delta^{\frac{1}{3}} \biggl[ e^{-\frac{2}{3} (4 U + V)} ds^2_{5} + e^{2 U} \frac{1- \tilde{c}y}{6} d s^2(S^2) 
\nonumber \\ && \ \qquad +  e^{2 U} \frac{dy^2}{w q} + \frac{e^{2 U + 2 V} wq}{9 \Delta} ( \hat{\sigma}^3 )^2 \biggr]
\nn && \ \qquad + \Delta^{-\frac{2}{3}} \left( e^{-\phi} (d \alpha + \tilde{B}_{(1)})^2 + e^{\phi} (d \beta + \tilde{C}_{(1)})^2 \right)  .
\eea
}For the four-form we find, after some calculation,
{\setlength\arraycolsep{0pt}
\bea
\label{upliftflux}  G^{(4)} &=&  - \frac{2}{9} e^Z (1-\tilde{c}y) dy \hat{\sigma}^3 \sigma^1 \sigma^{2} -  \frac{1}{3} K_2  dy \hat{\sigma}^3 \nn &+& \frac{1}{3} y (1-\tilde{c}y) K_2   \sigma^1 \sigma^{2} + e^{-\frac{4}{3} (U+V)} (* K_2 ) dy \nn &+& \frac{1}{3}(1-\tilde{c}y) K_1 dy  \sigma^1 \sigma^{2} -4 y e^{-8 U} (* K_1) \nn &+& \sqrt{\frac{1-\tilde{c}y}{6wq}} \biggl( L_2 (\sigma^2 - i \sigma^1)(2 y dy - i \frac{wq}{3} \hat{\sigma}^3  )  \nn &+& e^{-\frac{4}{3} (U+V)} (* L_2) i w q (\sigma^2 - i \sigma^1) + c.c. \biggr)  \nn
&+& d \tilde{B}_{(2)} (d \beta + \tilde{C}_{(1)}) \nonumber \\
&+& G_3 (d \alpha + \tilde{B}_{(1)}) + \frac{e^{2 U} wq}{3 \Delta} G_2 \hat{\sigma}^3  ( d \alpha + \tilde{B}_{(1)}) \nonumber \\
&-&  \frac{2 y e^{2 V}}{3 \Delta} dy  \hat{\sigma}^3 G_1  (d \alpha+ \tilde{B}_{(1)}) \nonumber \\ 
&+& \frac{1}{6} (1-\tilde{c}y) \sigma^1 \sigma^{2} G_1  (d \alpha + \tilde{B}_{(1)})  \nonumber \\ 
&-& \frac{2 y e^{2 V}}{3 \Delta} N_1 i w q \sqrt{\frac{1-\tilde{c}y}{6 w q}} (\sigma^1 + i \sigma^2) \hat{\sigma}^3 (d \alpha + \tilde{B}_{(1)})  \nonumber \\ 
&+&  N_1 \sqrt{\frac{1-\tilde{c}y}{6 wq}} (\sigma^2 - i \sigma^1) dy (d \alpha  +  \tilde{B}_{(1)}) \\
&+& N_0 \frac{e^{2U} w q}{ 3 \Delta} \sqrt{\frac{1-\tilde{c}y}{6 w q}} ( \sigma^2 - i \sigma^1) dy \hat{\sigma}^3 ( d \alpha + \tilde{B}_{(1)}), \nonumber
 \eea
}where we have omitted wedge products and $K_i, L_i, G_i, N_i$ refer to the fields in the original notation of \cite{oscar1}.

Several remarks about the consistent embedding (\ref{upliftmet}), (\ref{upliftflux}) are now in order. At generic points in the $N=4$ moduli space, the metric on the internal $M_6$ receives deformations from $\Delta$ that nevertheless preserve the cohomogeneity-one foliation by $y$ of the undeformed background metric. The $SU(2)\times U(1)^3$ symmetry of the background  is, however, generically broken by $\tilde{B}_{(1)}$, $\tilde{C}_{(1)}$ to $SU(2)\times U(1)$. As in the reduction to minimal supergravity of \cite{m6red}, the $U(1)$ symmetry associated to $\sigma^3$ is gauged by the $N=2$ graviphoton $A_1$. The present metric additionally contains the vectors $B_1$, $C_1$, which enter (\ref{upliftmet}) in an unconventional way, as they do not gauge any isometry of $M_6$ (they are involved in the gauging of a Heisenberg symmetry coming from the four-form). $B_1$, $C_1$  become massive after gauge-fixing the shift symmetries corresponding to the axions $b$, $c$ (see \cite{Cassani:2010uw, oscar1,Liu:2010sa}). That a metric produces massive vectors is not surprising as, in fact, an infinite tower of those arise upon compactification of the metric. What seems remarkable is that a finite number of massive vector metric modes can be retained in a full non-linear embedding like (\ref{upliftmet}). We are not aware of a similar case having been previously discussed in the literature.

From a purely $D=11$ perspective, the consistency of the truncation is quite  non-trivial. While the singlet and constant $SU(2)$ torsion classes of $SE_5$ guarantee the consistency of the type IIB truncation, one of the effects of the T-duality is to generate non-trivial modules in the $SU(2)$-structure of the $D=11$ background (see section \ref{geometries}), thus preventing the $D=11$ truncation to follow from any obvious symmetry principle. This is reflected by the  non-obvious manner in which the $D=5$ fields enter the $D=11$ ones. As a cross-check, we have partially verified that the $D=11$ field equations, evaluated on (\ref{upliftmet}), (\ref{upliftflux}) reproduce, as they should, the $D=5$ equations of motion (collected {\it e.g.} in appendix B of \cite{oscar1}). But, had not we obtained the full non-linear $D=11$ ansatz by T-dualising the simpler $SE_5$ ansatz (\ref{KKmet}), (\ref{KKforms}), it would have been extremely difficult to figure out what the $D=11$ KK ansatz would have been, or whether an ansatz from $D=11$ would exist at all.

Finally, recall that the $D=5$ $N=4$ theory admits a truncation to minimal $D=5$ $N=2$ supergravity (see again, {\it e.g.}, \cite{oscar1} for the details). The reduction ansatz (\ref{upliftmet}), (\ref{upliftflux}) then reduces to the one of \cite{m6red}, reviewed in section \ref{geometries} above, particularised to the case of $S^2 \times T^2$ base in $M_6$. The reduction of \cite{m6red} is, however, general for all the GMSW geometries $M_6$ of \cite{ads5m6} so a natural question is to ask whether the $D=5$ $N=4$ theory, or any subtruncation thereof, can be obtained from more general geometries within the GMSW class. Apart from the (singular) $H^2 \times T^2$ case, for which the truncation may be adapted to produce an $N=4$ supergravity, inspection of (\ref{upliftmet}) reveals that no supersymmetric subtruncation of the $N=4$ theory of \cite{Cassani:2010uw, oscar1}, other than the minimal $N=2$ one, can be obtained from a generic GMSW geometry $M_6$.
To see this, recall that the $N=4$ theory can be truncated to $N=2$ supergravity coupled to a universal hypermultiplet $(\phi, a, \xi, \xi^*)$. This is the minimal supersymmetric extension of $D=5$ minimal supergravity, that still is a subtruncation of the $N=4$ theory. From (\ref{upliftmet}), the effect of the scalars $(\phi, a, \xi, \xi^*)$ can be seen to deform the $M_4 = S^2 \times T^2$ base of $M_6$. Assuming that a truncation on generic $M_6$ should preserve the geometry of $M_4$, we reach the conclusion above.

What is conceivable is that consistent truncations to other matter-coupled theories in $D=5$ can be obtained from a generic GMSW geometry, by judiciously constructing a KK ansatz from the $SU(2)$-structure forms. Although we have not succeeded in building new truncations, in section \ref{Remarks} we comment on possible strategies to do this.

\section{New non-supersymmetric $AdS_5$ solutions} \label{solutions}

The original $AdS_5$ background is obviously recovered when all the $D=5$ excitations are turned off. In other words, the $N=2$ $AdS_5$ critical point of the $N=4$ theory uplifts to $D=11$ via (\ref{upliftmet}), (\ref{upliftflux}). Of course, this is guaranteed by the consistency of the truncation, as is the consistent uplift of any other solution to the $D=5$ theory. In particular, the $D=5$ potential admits a second, non-supersymmetric $AdS_5$ critical locus (first found by Romans \cite{Romans:1984an} in $D=5$ $N=8$ supergravity), at
\be \label{RomansCP}
e^{4U}=e^{-4V}=\frac{2}{3},\; \xi=\frac{1}{\sqrt{12}}e^{\frac{\phi}{2}}e^{i\theta^\prime},\; \chi-a\xi=ie^{-\phi}\xi ,
\ee
where $\theta^\prime$ is an arbitrary phase. All other scalars of the $D=5$ theory are set to trivial, and the $AdS_5$ radius
at these points is $2{\sqrt 2}/3$.

Feeding (\ref{RomansCP}) into (\ref{upliftmet}), (\ref{upliftflux}), we find two new classes of $D=11$ solutions, consisting on non-supersymmetric, warped and direct products, respectively, of $AdS_5$ with a smooth, complex manifold $M_6$, equipped, in both cases, with a Hermitian, two-parameter metric. Both classes are parametrised by $(\tilde{a}, \phi)$. To see this, note that out of the five parameters $(\tilde{a},\tilde{c}, a, \phi, \theta^\prime)$ of the solution (\ref{RomansCP}), (\ref{upliftmet}), (\ref{upliftflux}), $\theta^\prime$ and $a$ can be removed by coordinate transformations. The two classes are then distinguished by the value of $\tilde c$. If $\tilde{c} \neq 0$, it is easy to check that it can be taken to $\tilde{c}=1$ with no loss of generality, leading to the two-parameter warped product class. If $\tilde{c}=0$, the warp factor $\Delta$ reduces to a constant, leading to the direct product class. The non-supersymmetry of the $D=5$ critical point (\ref{RomansCP}) carries over to its $D=11$ uplift. This lack of supersymmetry can also be directly seen in $D=11$, as the uplifted solutions do not fit into the generic form of $AdS_5$ supersymmetric solutions in $D=11$ \cite{ads5m6}, reviewed in subsection \ref{geometries}.

The topology of the internal $M_6$ for both classes is the same as that of the supersymmetric $S^2 \times T^2$ solutions of \cite{ads5m6}, namely, a smooth trivial $T^2$ bundle over a four dimensional base $B_4$, which is itself a trivial $S^2$ bundle over $S^2$. Now, unlike the solutions in \cite{ads5m6}, the metric on $M_6$ is not a warped product of $T^2$ with $B_4 = S^2\times S^2$, but the former is (trivially) fibered over the latter by 
$\tilde{B}_{(1)} = -\frac{\sqrt{2}}{6} e^{-3\phi/2} \sqrt{(1-\tilde{c}y) w q } \ \sigma^1$
and 
$\tilde{C}_{(1)} = -\frac{\sqrt{2}}{6} e^{3\phi/2} \sqrt{(1-\tilde{c}y) w q } \ \sigma^2$. The topological triviality and metric regularity of $B_4$ follows from a similar analysis as that of \cite{ads5m6,ypq}, so here we will only show the triviality of the $T^2$ bundle. To see that $T^2$ does not wind over $B_4= S^2 \times S^2$, we first follow  \cite{ypq} closely and construct a basis $C_1, C_2$ in homology $H_2(B_4)$ out of the two copies of the round $S^2$, with coordinates $(\theta, \varphi$), located at the poles of the $S^2$ fiber $(y,\psi)$. These poles are themselves located at the two smaller roots $y_1$, $y_2$ of the numerator of $q(y)$ in (\ref{funwq}). It is easy to check that the curvatures $d\tilde{B}_{(1)}$,  $d\tilde{C}_{(1)}$ integrate to zero over the cycles $C_1$, $C_2$, thus showing the vanishing of the corresponding Chern numbers.

For both classes of solutions, the internal $M_6$ is complex. Introducing the obvious frame $e^i$, $i=1 ,\ldots, 6$, for the internal metric in (\ref{upliftmet}), (\ref{RomansCP}), it can be shown that the $(3,0)$-form $\Omega = (e^1 +i e^2) \wedge (e^3 +i e^4) \wedge (e^5 +i e^6)$ is such that $d \Omega = A \wedge \Omega$ for a suitable one-form $A$, different for each class. As a consequence, the associated almost complex structure is integrable and $M_6$, therefore, complex. $M_6$ is, however, not K\"ahler. Neither it is Einstein so, in particular, our direct product solutions are not in the class of non-supersymmetric $AdS_5 \times KE_6$ solutions \cite{Dolan:1984hv,Pope:1988xj}, where $KE_6$ is a K\"ahler-Einstein six-fold. A $D=5$ reduction related to this class of solutions was constructed in \cite{O'Colgain:2009yd}.

\section{Possible extension to general $M_6$}  \label{Remarks}

We now turn to the question of whether truncations beyond the one to minimal supergravity \cite{m6red} exist for other GMSW geometries \cite{ads5m6}. We start by considering generalisations to the explicit example $M_4 = S^2 \times S^2$, before commenting on the R-charged sector of the general GMSW solutions and the r\^{o}le of the warp factor in constraining potential reductions. As a general comment, note that (some combination of) the scalars $U$, $V$ might still be promoted to a reduction on general GMSW via the combination $\Delta$: using $2 y = e^{3 \lambda} \sin \zeta$ , equation (\ref{eq:delta}) becomes
\be
\label{DeltaUV}
\Delta = e^{6 \lambda}( e^{2 U} \cos^2 \zeta + e^{2 V} \sin^2 \zeta).
\ee

\subsection{KK reduction on $M_4= S^2 \times S^2$} \label{S2S2}
Neglecting $M_4 = H^2 \times T^2$, where the solutions are singular, the next most interesting explicit GMSW example is $M_4 = S^2 \times S^2$, a product of two-spheres that also serves as an example of $KE_4$ when the two spheres have equal radii. In order to consider any generalisation to this class of explicit solutions, a prerequisite is that one can find a reduction incorporating a single breathing mode.

Although the KK ansatz for $M_4 = S^2 \times T^2$ may be consistently truncated to a gravity scalar theory through setting all fields to zero and $U=V$, (see \cite{Bremer:1998zp,Liu:2000gk}), we have checked that such a reduction may not be generalised to $M_4 = S^2 \times S^2$. The failure of this simple reduction suggests that one must introduce an additional flux term in tandem with the breathing mode. Observe that for all the explicit solutions where $M_6$ is complex, the four-form flux may be expressed schematically as
$G^{(4)} = \frac{\tilde{c}}{6} \vol_4 + d [f(y) J \w D \psi ]$, thus immediately satisfying the Bianchi identity. The bracket under $d$ is a natural place to introduce a scalar into the four-form flux, in a compatible way with the Bianchi. Indeed, we are aware of similar construction \cite{Gauntlett:2007sm} of the reduction ansatz of $D=11$ supergravity on the Lin-Lunin-Maldacena (LLM) geometries \cite{LLM}  down to $D=5$ $SU(2) \times U(1)$ gauged supergravity \cite{romans}.

Despite the obvious difference in supersymmetry of the original backgrounds, one striking similarity is that both the LLM  geometries and $M_4 = S^2 \times T^2$ GMSW class involve $U(1)$ fibrations over a \textit{single} Riemann surface. When both the Riemann surfaces, or alternatively $KE_4$ is fibred, it is clear that the ansatz provided by studying $S^2 \times T^2$ will not be appropriate. The identification of a good candidate for the reduced gauged supergravity would be helpful, as it is easier to find the correct ansatz when one has a target reduced theory in mind. 

In the absence of a breathing mode, another candidate field that may be retained is $K_1$, the existence of which ensures $K_2$ and $F_2$ are not directly related. In fact, it is easy to retain $K_1$ so that the  Bianchi holds by essentially reversing a sign appearing in the reduction of \cite{m6red}. However, one finds that such an ansatz fails for $M_4 = S^2 \times S^2$. This is not completely unexpected, as we have dropped the breathing modes that play an important role in consistency when $M_4 = S^2 \times T^2$.

\subsection{R-charged fields} \label{Rcharges}

In the reduction of section \ref{Mtheory}, $D=5$ R-charged modes descending from both the $D=11$ metric and the four-form were produced from deformations of the one-form $\sigma^2 - i \sigma^1$ on $S^2$ which, together with its complex conjugate, transforms as a doublet of the $U(1)$ R-symmetry generated by $\sigma^3$. For general GMSW geometries, where there is a $U(1)$ fibration over the entire base, the natural R-charged object to be considered is the complex two--form $\Omega$ (see subsection \ref{geometries}).

A natural addition to the four-form reduction ansatz (\ref{fluxred}) is thus given by the combination 
\be
\label{closed} L_2 \wedge e^{6 \lambda} \Omega \cos \zeta + M_1 \wedge 3 e^{6 \lambda} \Omega \wedge (-K^1 \sin \zeta + i e^6),
\ee
together with its complex conjugate. With the help of the $M_6$ torsion condition \cite{ads5m6}
\bea
\label{susy3} e^{-6 \lambda} d (e^{6 \lambda} \Omega \cos \zeta) = 3 \Omega \w (-K^1 \sin \zeta + i K^2),
\eea
the combination (\ref{closed}) can be shown to be closed, and therefore compatible with the $D=11$ Bianchi identity, provided the complex spacetime forms $L_2$, $M_1$, satisfy 
\begin{equation} \label{BianchiLM}
DL_2 - iM_1 \wedge F_2 = 0 , \quad DM_1 + L_2=0 ,
\end{equation}
with $D L_2 \equiv dL_2 -i A_1 \wedge L_2$, and similarly for $DM_1$. Recall also that $F_2=dA_1$. Equation (\ref{BianchiLM}) shows that $L_2$ and $M_1$ are, respectively, field strengths for a $U(1)$ R-symmetry doublet of vectors and scalars.

If one imposes the restrictions $M_1 \wedge M_1^* =0$, $L_2 \wedge L_2^* = 0$, $M_1 \wedge L_2 = M_1 \wedge L_2^* = 0$ (which already signal that $D=5$ degrees of freedom are being neglected in (\ref{closed})), the equation of motion for the four-form  (\ref{fluxred}), (\ref{closed}) is also satisfied 
provided
{\setlength\arraycolsep{-3pt}
\begin{eqnarray} \label{eomLM}
&& D*L_2 -15 *M_1 + \frac{1}{3} F_2 \wedge L_2 -i M_1 \wedge *F_2 =0, \nonumber \\ 
&& D*M_1=0 .
\end{eqnarray}
}These equations of motion show that, in the $AdS_5$ vacuum, the $M_1$ scalars are St\"uckelberged away, giving a mass  $m^2= 15$ (in units of the inverse $AdS_5$ radius) to the vectors with field strength $L_2$. Unfortunately, even with the restrictions mentioned above, the ansatz (\ref{fluxred}), (\ref{closed}) proves inconsistent at the level of the $D=11$ Einstein equation. 

This is primarily due to the fact that all the deformations of the original GMSW flux, namely (\ref{fluxred}) and (\ref{upliftflux}), appear in orthonormal frame with the factor $e^{-\lambda}$. This requirement comes from the Einstein equations where the overall metric warp factor $e^{2 \lambda}$ appears inverted, forcing the $G^2$ terms in orthonormal frame to appear with $e^{-2 \lambda}$ factors. Neglecting the generalised calibration (see \cite{m6red}) which allows $J$ in (\ref{fluxred}) to appear with the correct factor, this explains the absence of other terms involving $J$ and $\Omega$. One can add other terms to the four-form, like (\ref{closed}) wedged with $e^{3 \lambda} \cos \zeta K^1$, but similar inconsistencies end up arising.

Interestingly, vectors with $m^2 = 15$ also appear in the reduction of type IIB supergravity on $S^5$ \cite{Kim:1985ez}. It is tempting to speculate that these are related to ours, as $S^5$, like the $Y^{p,q}$ metrics discussed above, can be recovered from $M_6$ by reduction and T-duality in the case of $S^2 \times T^2$ base.

\section{Discussion} \label{Conclusions}

The truncation we presented in section \ref{Mtheory} is one of the few examples of a consistent truncation on a non-trivial $G$-structure manifold to matter-coupled supergravity, that is not apparently driven by symmetry principles. Another example we are aware of in this class is the truncation of $D=11$ supergravity on a certain $SU(2)$-structure seven-fold \cite{Gauntlett:2006ux} down to $D=4$ $N=2$ gauged supergravity coupled to a vector multiplet and two hypermultiplets \cite{Donos:2010ax}.

In both these two cases, the reduction ansatz was proposed by indirect methods. Here, from T-duality of a type IIB ansatz whose consistency, before T-dualisation, does stem from symmetry principles. In \cite{Donos:2010ax}, by first reducing to $D=7$ $SO(5)$--gauged supergravity \cite{Nastase:1999cb}, and then further reducing on three dimensions whilst keeping $SO(3)$ singlets in the decomposition $SO(5) \rightarrow SO(3) \times SO(2)$. From the $D=11$ perspective, however, both are matter-coupled truncations on non-trivial $G$-structure manifolds, that do not follow from any obvious symmetry principle. Here, non-trivial torsion classes in the internal geometry are generated by T-duality; in \cite{Donos:2010ax}, the $D=11 \rightarrow D=7$ step does not follow from symmetry, and this accordingly translates into the straight $D=11 \rightarrow D=4$  reduction. For other examples of consistent KK reductions obtained from T-duality, see \cite{Cvetic:2000yp}.

This note has also made apparent how difficult it is to construct reductions when there are no guiding symmetry principles to write a KK ansatz out of the internal $G$-structure, or when the target lower-dimensional theory is unknown. In section \ref{Remarks}, we have commented on this kind of difficulties for reductions on generic GMSW geometries \cite{m6red}. 
Inconsistencies typically arise when the warp factor $\lambda$ does not drop from the $D=5$ equations. It would be interesting to determine if the inclusion of a moduli-dependent warp factor, such as (\ref{DeltaUV}), introduced into the KK ansatz using the hints in subsection \ref{S2S2}, would restore consistency.

Using the embedding of section \ref{Mtheory}, we have constructed two new classes of smooth, non-supersymmetric $M$-theory backgrounds containing, respectively, warped and direct products of $AdS_5$ with a compact space $M_6$. We have built these solutions via $D=11$ uplift of the non-supersymmetric critical point \cite{Romans:1984an} of the $D=5$ $N=4$ theory of \cite{Cassani:2010uw,oscar1}. The type IIB uplift of this point can be expected to be unstable \cite{Bobev:2010ib}, so it would be interesting to determine if instabilities also arise for our $D=11$ solutions. Observe, however, that the non-supersymmetric solution $AdS_5 \times CP^3$ \cite{Dolan:1984hv,Pope:1988xj} is known to be stable \cite{Martin:2008pf}.

Similar supersymmetric $AdS_3 \times M_8$ solutions to GMSW exist \cite{ads3m8} where $M_8$ is an $S^2$-bundle over $M_6=KE_6$ or products of K\"{a}hler-Einstein spaces. In accordance with the conjecture of \cite{Gauntlett:2007ma}, one expects a consistent truncation retaining an R-symmetry vector. Moreover, our findings here suggest that when $M_6$ has a $T^2$ factor, a more general reduction to $D=3$ supergravity, such as that in \cite{Colgain:2010rg}, may be found.

More generally, it would be very interesting to determine the conditions that allow for consistent matter couplings in KK reductions on non-trivial $G$-structure manifolds.

\section*{Acknowledgements}	
We have enjoyed conversations with Changhyun Ahn, Jerome Gauntlett, Nakwoo Kim, Tetsuji Kimura, Tristan McLoughlin and Hossein Yavartanoo. OV wishes to thank KIAS for kind hospitality during the early stages of this work. OV is partially supported by the Spanish Government research grant FIS2008-01980.

\end{document}